\documentclass[12pt]{article}
\usepackage[left=1in,top=1in,right=1in,bottom=1in,nohead]{geometry}
\usepackage{amsmath}
\usepackage{amssymb}
\usepackage{bm}
\usepackage{lineno}
\RequirePackage[OT1]{fontenc}
\RequirePackage{amsthm,amsmath}
\RequirePackage[numbers]{natbib}
\RequirePackage[colorlinks,citecolor=blue,urlcolor=blue]{hyperref}
\usepackage{amsmath}
\usepackage{graphicx,psfrag,epsf}
\usepackage{enumerate}
\usepackage{url}

\usepackage{setspace}

\usepackage{bm}         
\usepackage{amsmath}    
\usepackage{amsfonts}   
\usepackage{amssymb}    
\usepackage{multirow}
\usepackage{morefloats}
\usepackage{subcaption}
\usepackage{caption} 
\usepackage{bm}         
\usepackage{amsmath}    
\usepackage{amsfonts}   
\usepackage{amssymb}
\usepackage{natbib}
\onehalfspace 
\numberwithin{equation}{section}
\theoremstyle{plain}

\usepackage{mathtools}

\usepackage[figuresright]{rotating}
\usepackage{hyperref}
\newcommand{\footremember}[2]{%
	\footnote{#2}
	\newcounter{#1}
	\setcounter{#1}{\value{footnote}}%
}

\title{Change-point analysis in Financial Networks}
\author{%
	Sayantan Banerjee$^1$\footremember{alley}{Corresponding author at OM \& QT Area, IIM Indore, Indore, M.P. 453556, India. e-mail: sayantanb@iimidr.ac.in}%
	\and Kousik Guhathakurta$^2$
}
\date{%
	$^1$Operations Management \& Quantitative Techniques Area, Indian Institute of Management Indore\\%
	$^2$Finance \& Accounting Area, Indian Institute of Management Indore\\[2ex]%
}

\modulolinenumbers[50]

\begin{document}

\maketitle

\begin{abstract}

A major impact of globalization has been the information flow across the financial markets rendering
them vulnerable to financial contagion. Research has focused on network analysis techniques to
understand the extent and nature of such information flow. It is now an established fact that a stock
market crash in one country can have a serious impact on other markets across the globe. It follows
that such crashes or critical regimes will affect the network dynamics of the global financial markets.
In this paper, we use sequential change point detection in dynamic networks to detect changes in the network characteristics of thirteen stock markets across the globe. Our method helps us to detect changes in network behavior across all known stock market crashes during the period of study. In most of the cases, we can detect a change in the network characteristics prior to crash. Our work thus opens the possibility of using this technique to create a warning bell for critical regimes in financial markets.
\vspace{0.1in}

\end{abstract}

\section{Introduction}
Integration of global markets have resulted in higher interdependence between the financial markets. One emergent method of capturing the interdependence has been network analysis \cite{nobi2014correlation,zhao2016structure,bhattacharjee2017network,wang2018correlation}. Critical events like stock market crashes are likely to impact stock markets across the globe \cite{bertero1990structure,aloui2011global,kaue2012structure,luchtenberg20152008}. 

The credit crisis of 2008 has propelled a volume of research on stock market interdependence and the resulting contagion. Kenourgios et al. \cite{kenourgios2007financial}, using a multivariate regime-switching copula model found evidence of contagion effect between BRIC countries and US and UK stock markets during five financial crises. Markwat et al. \cite{markwat2009contagion} found `domino effect' across global stock markets where a crash in a local market has impact on multiple stock markets. Boubaker et al. \cite{boubaker2016financial} confirmed contagion effect between the stock markets of the US and some selected developed and emerging countries during the sub-prime crisis. Longstaff \cite{longstaff2010subprime} investigated sub-prime asset pricing during the sub-prime crisis and established contagion effect across markets. Proposing a wavelet-based approach to test for financial contagion, Galllegati \cite{gallegati2012wavelet} concluded that global stock markets were impacted by the US sub-prime crisis.  These and many other works \cite{jin2016global,bekaert2011global,bekaert2014global} have shown that critical periods in stock price movement like stock market crashes always impact multiple markets across the globe. This means that the nature of the interdependence of global stock markets undergo changes during such periods. Both research and practice in stock markets need to focus on methods of modelling this interdependence and devise methodology to detect change points in the time evolution of such interdependence. Mantegna \cite{mantegna1999hierarchical} provided one of the earlier works in exploring the dynamics among various stocks by constructing a minimal spanning tree (MST) using correlation between the stocks. Onnela et al. \cite{onnela2003dynamic} explored the dynamics of the MST constructed from the correlation matrix of stock returns to detect market crises. Mizuno et al. \cite{mizuno2006correlation} also studied correlation networks among currencies to construct a currency MST tree followed by clustering of the currencies. In a recent work, Kazemilari and Djauhari \cite{kazemilari2015correlation} analyzed the topological properties of stocks using the RV coefficient corresponding to multivariate time series data. Namaki et al. \cite{namaki2011network} used random matrix theory to analyze stock market network and provided some useful insights on the interdependence of individual stock prices in US and Tehran stock markets.

An interesting problem in this regard is that of change-point detection. Since we are considering networks which evolve over time, structural changes occur in the networks as the dynamics among the players of the network change significantly. These changes might be attributed to external factors affecting some or all of the players involved. Change-point detection has been widely studied in the context of statistical process control; see \cite{zamba2006multivariate,lowry1992multivariate} and references therein. McCulloh and Carley \cite{mcculloh2011detecting} used concepts from process control for detecting change-points in social networks varying over time. Aue et al. \cite{aue2009break} considered the problem in the context of low dimensional multivariate time series and developed a nonparametric test statistic for detecting change point in the coviance structure. Barnett and Onnela \cite{barnett2016change} used a testing based approach as well using correlation networks for detecting multiple change points. Moving to the high dimensional set up, Roy et al. \cite{roy2017change} developed a two-step algorithmic approach for single change point detection in networks corresponding to sparse Markov random fields. More recently, Avanesov and Buzun \cite{avanesov2018change} developed a change point detection method in high dimensional covariance structure of the underlying variables using statistical hypothesis testing. Peel and Clauset \cite{peel2015detecting} adopted a testing of hypothesis approach as well to infer whether a change point  has occured in a certain window of time, using a parametric family of distributions for the data. They used the concept of Bayes factor, and determined the change point using a $p$-value approach, similar to concepts in \cite{basseville1993detection}. Gaussian graphical models (GGMs), where the underlying random variables defined over the nodes of the network are assumed to be normally distributed, provide an utilatarian statistical framework for modeling dynamic networks \cite{zhou2010time}. Kolar and Xing \cite{kolar2011time} consider the estimation problem of such networks, followed by another work where they developed the method for estimating networks with jumps using a varying coefficient varying structure (VCVS) model \cite{kolar2012estimating}. In a recent work, Keshavarz et al. \cite{keshavarz2018sequential} considered a sequential change point detection algorithm for time varying sparse GGMs. A nonparametric sequential alorithm for detecting sudden changes in a network has been developed by Park and Sohn \cite{park2018detecting}.

We consider the problem of sequential change point detection in a dynamic network of a group of international stock markets to detect changes in the network dynamics during periods close to stock market crashes. Our work focuses on capturing the dynamics of the financial market via construction of the underlying network structure of the Asia -- Pacific stock markets, which present an interesting combination of emerging and developed markets, along with US and UK stock markets using stock return data from April 2000 to April 2017. We use the Gaussian kernel function which provides a distribution free construction of underlying network while simultaneously preserving the topology. This gets rid of the assumption about the underlying variables being multivariate normal which for financial returns might not hold true in general. This is followed by a statistical testing approach for change point detection in time-evolving networks as proposed in \cite{barnett2016change}. 

This work contributes to the literature in two-fold ways. First, we bring methodological innovation in network analysis using Gaussian kernel function, which renders the model more suited for non-normal variables like financial returns. With the adoption of Gaussian radial basis kernel in the change detection method proposed by \cite{barnett2016change}, we extend their model and bring in more efficiency in detecting change points. Additionally, we apply this method to financial returns of a group of stock markets of strategic importance and detect the impact of local crashes across markets. This opens the possibility of using our method as a warning signal. Moreover, we also illustrate that the kernel based network performs superior compared to the correlation network in terms of detecting valid change points.

\section{Data description}
We consider daily data for 11 stock market indices from Asia Pacific region: Australia (AORD), Hong Kong (HSI), Japan (NIKKEI), Singapore (STI), China (SSE), India (NIFTY), Indonesia (JKSE), South Korea (KOSPI), Malaysia (KLSE), Philippines (PSEI) and Taiwan (TWII), along with the US (DJI) and the UK (FTSE) stock market indices, for the period of April 2000 to April 2017. The data was taken from Thomson Reuters database. 

\begin{figure}[h]
	\centering
	\includegraphics[scale = 0.3]{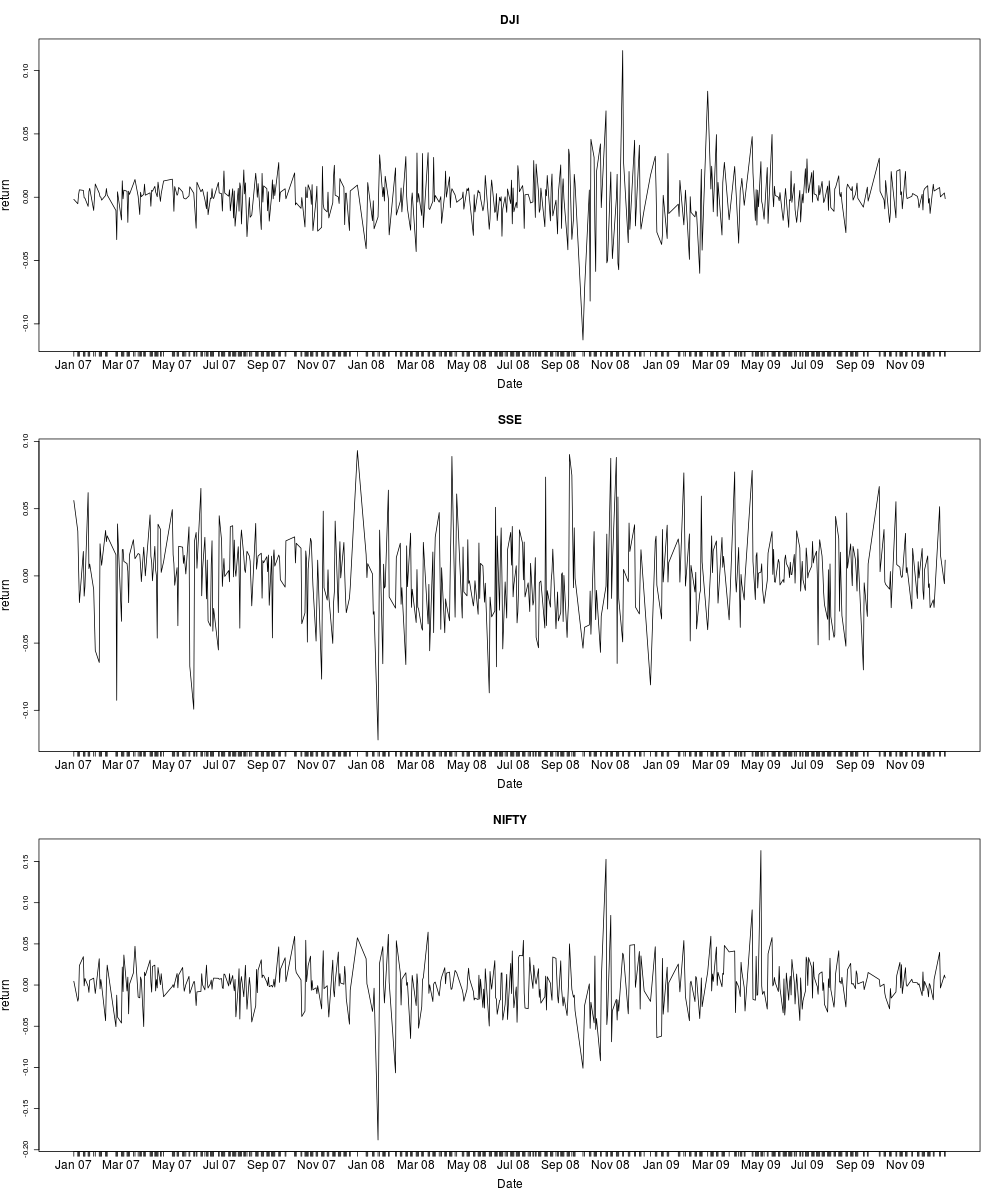}
	\caption{Figure showing plots of stock returns for three indices -- DJI, SSE and NIFTY, during the period of January 2007 to January 2009. This timeline includes the 2007-08 global financial crisis period. The patterns in the returns for the different stock exchanges are different, prompting the need for a network based analysis to explore the dynamics of global stock markets.}
	\label{fig:stockplot}
\end{figure}

In Figure~\ref{fig:stockplot}, we plot the stock market returns corresponding to the indices -- DJI, SSE and NIFTY, for the time period of January 2007 to January 2009. This timeline includes the 2007-08 global financial crisis period. Network based analysis will help in capturing the interplay among the various indices and explain the changing dynamics of the network before or after stock market crashes.

We construct the network based on the daily returns computed as the first difference of the level data. We observe that the following stock market crashes took place during the period.

\begin{enumerate}
	\item {\textbf{11-Sep-01, Economic effects arising from the September 11 attacks, USA.}}
	
	The September 11 attacks caused global stock markets to drop sharply. The attacks themselves caused approximately \$40 billion in insurance losses, making it one of the largest insured events ever. 
	
	\item {\textbf{09-Oct-02, Stock market downturn of 2002, USA.}}
	
	A downturn in stock prices during 2002 was observed in stock exchanges across the United States, Canada, Asia, and Europe. After recovering from lows reached following the September 11 attacks, indices slid steadily starting in March 2002, with dramatic declines in July and September leading to lows last reached in 1997 and 1998.
	
	\item {\textbf{27-Feb-07, Chinese Stock Bubble of 2007, China.}}
	
	The SSE Composite Index of the Shanghai Stock Exchange tumbled 9\% from unexpected selloffs, the largest drop in 10 years, triggering major drops in worldwide stock markets.
	
	\item {\textbf{11-Oct-07, United States bear market of 2007 - 09, USA.}}
	
	Till June 2009, the Dow Jones Industrial Average, Nasdaq Composite and S\&P 500 all experienced declines of greater than 20\% from their peaks in late 2007.
	
	\item \textbf{16-Sep-08, Financial crisis of 2007 - 08}
	
	On September 16, 2008, failures of large financial institutions in the United States, primarily due to exposure of securities of packaged subprime loans and credit default swaps issued to insure these loans and their issuers, rapidly devolved into a global crisis resulting in a number of bank failures in Europe and sharp reductions in the value of equities (stock) and commodities worldwide. The failure of banks in Iceland resulted in a devaluation of the Icelandic kr{\'{o}}na and threatened the government with bankruptcy. Iceland was able to secure an emergency loan from the IMF in November. Later on, U.S. President George W. Bush signed the Emergency Economic Stabilization Act into law, creating a Troubled Asset Relief Program (TARP) to purchase failing bank assets. This had disastrous effects on the world economy along with world trade.
	
	\item {\textbf{27-Nov-09, 2009 Dubai debt standstill, United Arab Emirates.}}
	
	Dubai requested a debt deferment following its massive renovation and development projects, as well as the Great Recession. The announcement caused global stock markets to drop.

	\item {\textbf{27-Apr-10, European sovereign debt crisis, Europe.}}
	
	Standard \& Poor's downgraded Greece's sovereign credit rating to junk four days after the activation of a 45-billion Euros EU-IMF bailout, triggering the decline of stock markets worldwide and of the Euro's value, and furthering a European sovereign debt crisis.
	
	\item {\textbf{06-May-10, 2010 Flash Crash, USA.}}
	
	The Dow Jones Industrial Average suffered its worst intra-day point loss, dropping nearly 1,000 points before partially recovering.
	
	\item {\textbf{01-Aug-11, August 2011 stock markets fall, USA.}}
	
	Stock markets around the world plummeted during late July and early August, and remained volatile for the rest of the year.
	
	\item {\textbf{12-Jun-15, 2015 -- 16 Chinese stock market crash, China.}}
	
	China stock market crash started in June and continued into July and August. In January 2016, Chinese stock market experienced a steep sell-off which sets off a global rout.
	
	\item {\textbf{18-Aug-15, 2015 -- 16 stock market selloff, United States.}}
	
	The Dow Jones fell 588 points during a two-day period, 1300 points from August 18 -- 21. On Monday, August 24, world stock markets were down substantially, wiping out all gains made in 2015, with interlinked drops in commodities such as oil, which hit a six-year price low, copper, and most of Asian currencies, but the Japanese yen, losing value against the United States dollar. With this plunge, an estimated ten trillion dollars had been wiped off the books on global markets since June 3.
	
\end{enumerate}

\section{Network construction}
We have a fixed set of $n$ nodes for constructing the financial network corresponding to $n$ different stocks. This results in $T$ dimensional observations $Y_1,\ldots,Y_n$ at $T$ fixed time points $t = 1,\ldots,T$. Let $Y_i = (Y_{i1},\ldots,Y_{iT})^T$ denote the vector of observations for stock $i,\, i = 1,\ldots,n$, and the resulting $n \times T$ data matrix be denoted by $\mathbb{Y}.$ Based on the observed data matrix $\mathbb{Y}$, we aim to construct a similarity graph or network $G = (V,E)$, where $V = \{1,\ldots,n\}$ is the set of nodes corresponding to $\{Y_1,\ldots,Y_n\}$ and $E$ is the collection of edges. The resulting network may be an unweighted or a weighted one, depending on whether the `similarity' associated with any pair $(i,j)$ of edges is binary, or some positive quantity $w_{ij}$, respectively. We say that two nodes are connected with edge-strength $w_{ij}$ if $w_{ij} > 0.$ In this context, the weighted adjacency matrix of a network is defined as the symmetric $n \times n$ matrix $W = (\!(w_{ij})\!)$.

\subsection{Similarity Networks}
We now discuss different types of similarity networks that can be constructed in this context.

{\textbf{$\epsilon$-neighborhood networks:}} An $\epsilon$-neighborhood network is constructed by connecting the nodes for which the distance between two nodes exceeds a certain threshold $\epsilon > 0.$ The distance measure can be any well-defined metric, for example, the $L_1$ distance or the $L_2$ distance. The resulting network is an unweighted one.

{\textbf{Correlation/Partial Correlation networks:}} Correlation networks are constructed using the absolute value of the pairwise correlation between any two nodes as the corresponding edge-weights, that is, $w_{ij} = |r_{ij}|,$ $r_{ij}$ being the sample correlation coefficient between $Y_i$ and $Y_j$. Correlation networks have been extensively studied in various fields including analyzing protein networks (see \cite{zalesky2012use} and references therein), and financial networks \cite{barnett2016change,mizuno2006correlation,wang2018correlation}. Partial correlation networks are conceptually similar, with the main difference being the choice of the weight function as the partial correlation between pairs of nodes instead of correlations. Partial correlation between any two nodes adjust for the effect of the remainder of the nodes and hence might be more appropriate. Banerjee et al. \cite{banerjee2019spectral} used partial correlation graphs to model proteomic data in the context of graph clustering.

{\textbf{Kernel based networks:}} Kernel-based networks result in a fully connected graph where the underlying similarity measure captures the local neighborhood relationships of the network. The kernel function acts as a similarity measure between any pair of nodes, such that similar behaving nodes have higher edge weights. For example, for a Gaussian radial-basis kernel, the strength of an edge between two nodes $Y_i$ and $Y_{j}$ is given by
\begin{equation}
w_{ij} = \exp\left(-\dfrac{1}{2\sigma^2}\|Y_i - Y_{j}\|^2\right).
\end{equation}
The parameter $\sigma^2$ above controls the width of the node-neighborhoods in the network. Other choices of the kernel function include Laplacian kernel, polynomial kernel, Spline kernel, Bessel kernel, etc.

We shall be using the Gaussian kernel function to construct our financial network. Note that the network can be constructed using any sub-vector of $Y_i$ as well. One major reason behind the choice is that the network corresponding to the $n$ nodes is constructed under a non-parametric framework without any distributional assumptions on the columns of $\mathbb{Y}.$ In comparison, partial correlation networks effectively capture the conditional independence structure of the underlying variables in the nodes, but only under the assumption that the variables are multivariate Normal. Hence kernel-based approaches provide a distribution-free construction of the underlying network while simultaneously preserving the neighborhood topology.

\section{Change-point detection method}

We adopt the method proposed in \cite{barnett2016change} for multiple change point detection. While their method was developed for correlation networks, we shall be primarily working with the network constructed using the Gaussian radial basis kernel as described in the previous section. For two contiguous time windows $[t_1,t_2]$ and $[t_2+1,t_3]$ where $1 \leq t_1 < t_2 < t_2 + 1 < t_3 \leq T$, we say that a change-point exists at the boundary $t_2$ if the networks constructed using the respective time windows are significantly different. 

Let us define the adjacency matrix of the network constructed using the sub-vector of observations in the time range $[t,t']$ as $W(t,t')$. The presence of a change-point at the time point $t+k$ in a given time interval $[t,t'], \, t < t + k < t'$, should result in a different network topology in the respective time intervals $[t,t+k]$ and $[t+k+1,t'].$ This change in the topology may be captured using some suitable distance metric like the Euclidean norm, defined as
\begin{equation}
d(k) = \|W(t,t+k) - W(t+k+1,t')\|_2^2.
\end{equation}
If $\mathcal{W}(t,t+k)$ and $\mathcal{W}(t+k+1,t')$ respectively represent the `true' adjacency matrices corresponding to the network in the given time intervals, then our problem boils down to testing the hypothesis
\begin{equation*}
H_0: \mathcal{W}(t,t+k) = \mathcal{W}(t+k+1,t') \; vs.\; H_1: \mathcal{W}(t,t+k) \neq \mathcal{W}(t+k+1,t'),
\end{equation*}
for some $k \in (t,t').$ We propose to use the distance metric defined above as the test statistic. However, the exact sampling distribution of the same is not known. Hence we shall adopt a resampling approach to generate values of $d(k)$ under the assumption that the null hypothesis $H_0$ is true. For this, we first generate bootstrap samples of the rows of the data matrix corresponding to the observations in the time interval $[t,t']$. Let us denote the bootstrap resamples of the corresponding data matrix as $\mathbb{Y}^{(b)},\, b = 1,\ldots,B,$ where $B$ is the total number of bootstrap resamples. For each such data matrix, we construct the bootstrap resampled network using the Gaussian radial basis kernel function for the time intervals $[t,t+k]$ and $[t+k+1,t']$, where $k \in [t+\delta,t'-\delta],\, \delta > 0$. The quantity $\delta$ is chosen in such a way that we do not search for a potential change point near the end-points $t$ or $t'$. Thus we finally get resampled values of the test statistic as 
\begin{equation}
d^{(b)}(k) = \|W^{(b)}(t,t+k) - W^{(b)}(t+k+1,t')\|_2^2,
\end{equation}
where $W^{(b)}(t,t+k)$ and $W^{(b)}(t+k+1,t')$ respectively denote the weighted adjacency matrices in the time intervals $[t,t+k]$ and $[t+k+1,t']$ based on the resampled data matrix $\mathbb{Y}^{(b)}.$ A $z$-score corresponding to the potential change-point $k \in \{t+\delta, \ldots, t'-\delta\}$ is given by
\begin{equation}
z^{(b)}(k) = \dfrac{d^{(b)}(k) - \bar{d}(k)}{s(k)},
\end{equation}
where $\bar{d}(k) = \sum_{b=1}^{B}d^{(b)}(k)/B,$ and $s^2(k) = \sum_{b=1}^{B}\{d^{(b)}(k) - \bar{d}(k)\}^2/(B-1)$.

Note that, if autocorrelations are present among the observations, then resampling using an independent bootstrap as above might lead to a bias in approximating the null distribution. In that case, a sieve bootstrap may be used as outlined in \cite{barnett2016change}. 

A potential change-point is then identified as the time point having the maximum $z$-score, so that we define $Z^{(b)} = \max_k\{z^{(b)}(k)\}, b = 1,\ldots,B.$ We calculate a similar $z$-score for the observed data $\mathbb{Y}$, given by
\begin{equation}
z(k) = \dfrac{d(k) - \bar{d}(k)}{s(k)}.
\end{equation}
The resulting p-value of the test is thus given by
\begin{equation}
p = \dfrac{1}{B}\#\{b: Z^{(b)} \geq Z \},
\end{equation}
where $Z = \max_k z(k).$ If the resulting $p$-value is lesser than some chosen level of significance $0 < \alpha < 1$, then we reject the null hypothesis and conclude that a change point exists at the time point $k = \arg \max_k z(k).$

For detecting multiple change points over a longer interval, we can search for potential change points by splitting the interval into smaller intervals and then repeating the above method for each of the different segments. Apart from this, a chosen sub-interval might also have multiple change points. In that case, as discussed in \cite{barnett2016change}, after we have detected a certain change point at the time point $k \in [t,t']$, we can search for potential change points within the sub-intervals $[t,t+k]$ and $[t+k+1,t']$ individually, and repeat it further over smaller sub-intervals.

\section{Results}

We applied our change point detection method on different windows of length 200 days with choice of $\delta$ to be 50, so as to detect a change point $k \in \{t+50,t'-50\}$ for $|t'- t| = 200.$  We checked for the presence of autocorrelations among the daily stock returns using a Durbin-Watson testing procedure but didn't get sufficient evidence in favor of the same. This lead us to use independent bootstrap for resampling the rows of the data matrix.

\begin{figure}[h]
	\centering
	\includegraphics[scale = 0.5]{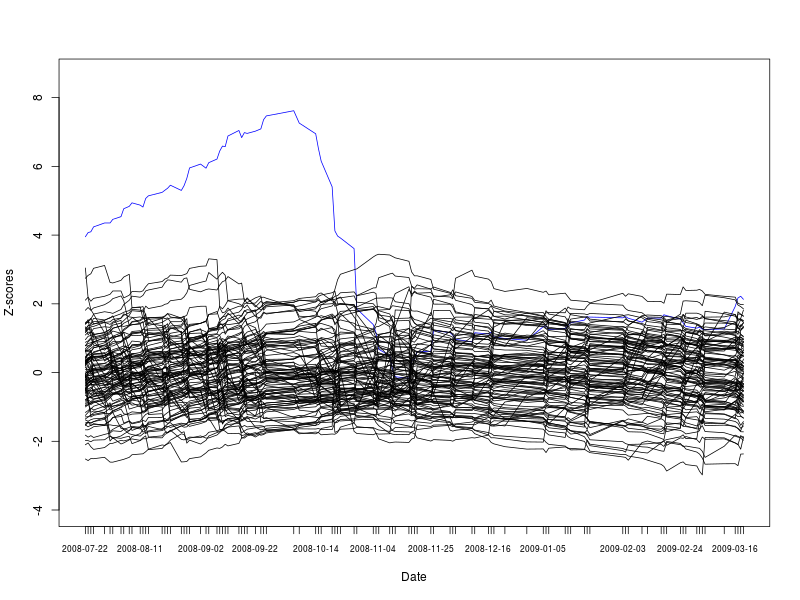}
	\caption{Figure showing plot of the z-scores for the observed data (blue line) against the z-scores corresponding to the bootstrapped samples (black lines), for networks analysed over a 200-day time period of 21-April-2008 to 30-June-2009.}
	\label{fig:chpointplot}
\end{figure}

Our change point analyses reveal that the dates where the network dynamics undergoes change, are all close to some stock market crash. Table~\ref{Table:Results} list the dates and the related stock market events. In Figure~\ref{fig:chpointplot}, we show the plots of the z-scores for the observed data (blue line) against the z-scores corresponding to the bootstrapped samples (black lines), for networks analysed over a 200-day time period of 21-April-2008 to 30-June-2009. Our method scanned for change points in a set of 101 contiguous dates, excluding $\delta = 50$ days on either side of the aforementioned dates. The true z-scores reach the peak on 06-October-2008, and has a $p$-value of zero, thus confirming a change-point in the network around that time period. The date corresponds to the global finanical crisis of 2007-08.

If we carefully observe the change points, we find that the cases where no change point were detected, that is, the 9/11 and 2010 Flash crash -- were exogenous crashes where stock markets took a temporary dip due to unforeseen circumstances. In all other cases, however, there were stock market bubbles building over a time and the shocks were endogenous in nature. The only exception is the Dubai 2009 debt standstill.  That may be expected as most of the major stock markets like FTSE shrugged off the effect of the Dubai crisis within a few days. A closer inspection reveals that the time gap between the change point and the reported crash dates is different for different events. That may be explained from the fact that the impact of different crashes on the network dynamics of multiple stock markets is not same for all events. In some cases, the stock markets of countries other than the place of origin of crash react with a lag while in others the spread of contagion is instantaneous. What is evident from our studies that most stock market crashes significantly impact multiple stock markets, albeit in varying degrees. Our method has detected the break in network dynamics in all the cases where there was a known contagion effect.

\begin{table}[h]
	\begin{tabular}{llll}
		\hline
		\hline
		{\textbf{Name}}                                                                                                & {\textbf{Date}}      & {\textbf{Country}}                                                           & {\textbf{Change Point}} \\ 		\hline 
		\begin{tabular}[c]{@{}l@{}}Economic 			effects arising from \\the September 11 attacks\end{tabular} & 11-Sep-01 & United States                                                             & \begin{tabular}[c]{@{}l@{}}No change\\ 			Point\end{tabular}          \\
		&&&\\
		\begin{tabular}[c]{@{}l@{}}Stock market 			downturn of 2002\end{tabular}                          & 09-Oct-02 & United States                                                              & 13-Nov-02                                                             \\
		&&&\\
		\begin{tabular}[c]{@{}l@{}}Chinese stock			bubble of 2007\end{tabular}                           & 27-Feb-07 & China                                                             & 27-Dec-06                                                             \\
		&&&\\
		\begin{tabular}[c]{@{}l@{}}United States 			bear market of 2007-09\end{tabular}                   & 11-Oct-07 & United States                                                     & 09-Oct-07                                                             \\
		&&&\\
		\begin{tabular}[c]{@{}l@{}}Financial			crisis of 2007-08\end{tabular}                            & 16-Sep-08 & United States                                                     & 06-Oct-08                                                             \\
		&&&\\
		\begin{tabular}[c]{@{}l@{}}2009 Dubai 			debt standstill\end{tabular}                             & 27-Nov-09 & \begin{tabular}[c]{@{}l@{}}United Arab\\ 			Emirates\end{tabular} & \begin{tabular}[c]{@{}l@{}}No change\\ 			point detected\end{tabular} \\
		&&&\\
		\begin{tabular}[c]{@{}l@{}}European 			sovereign debt crisis\end{tabular}                         & 27-Apr-10 & Europe                                                            & 09-Feb-2010                                                           \\
		&&&\\
		\begin{tabular}[c]{@{}l@{}}2010 Flash			Crash\end{tabular}                                       & 06-May-10 & United States                                                     & \begin{tabular}[c]{@{}l@{}}No change\\ 			point detected\end{tabular} \\
		&&&\\
		\begin{tabular}[c]{@{}l@{}}August 2011 			stock markets fall\end{tabular}                         & 01-Aug-11 &        USA                                                           & 27-Jul-11                                                             \\
		&&&\\
		\begin{tabular}[c]{@{}l@{}}2015-16 			Chinese stock market crash\end{tabular}                     & 12-Jun-15 & China                                                             & 16-Jun-15                                                             \\
		&&&\\
		\begin{tabular}[c]{@{}l@{}}2015-16			stock market selloff\end{tabular}                           & 18-Aug-15 & United States                                                     & 19-Aug-15        \\
		\hline                                                    
	\end{tabular}
	\caption{Results of the change-point detection method.}
	\label{Table:Results}
\end{table}

For comparison, we applied the change-point detection procedure for a correlation network as well. The method based on correlation network failed to detect almost all the major stock market crises, except for the financial crisis of 2007 - 2008 (change point detected on 12-August-2008). Additionally, it detected two change points on 28-July-2006 and 10-August-2016. With no reported stock market crash during this time it is very unlikely that the global stock market correlation underwent any significant change during that period.

\section{Discussion}
Literature has shown that crashes at local stock markets often impact multiple markets and spread across them. A natural corollary of this phenomenon is that the network dynamics of stock markets across the globe are likely to undergo critical shift. In this paper we create a network of Asia-Pacific stock markets along with US and UK stock markets using daily index data. We use the Gaussian kernel function which provides a distribution free construction of underlying network while simultaneously preserving the topology. To understand the impact of critical regimes we detect change points in the network adopting the method presented in \cite{barnett2016change} with a modification allowing for accommodation of Gaussian kernel. On inspection, the dates corresponding to the change points detected are found close to reported crash dates of any of the stock markets in the network, especially those crashes which were endogenous in nature. Our work contributes both to the methodology and application of network change point detection. On the methodological side we contribute by modifying the method presented in \cite{barnett2016change}. This allowed us to build the network with gaussian kernel which has its obvious advantages as explained above. On the application side the ability of our method to detect change points in case of all the known endogenous stock market crashes makes it useful in distinguishing endogenous and exogenous crashes. This may also serve as a warning bell for change in market dynamics for investors exposed to multiple stock markets across globe.

\section*{Acknowledgements}
S.B. is partially supported by DST INSPIRE Faculty Award, Govt. of India, Grant No. 04/2015/002165.

\bibliographystyle{apalike}
\bibliography{chpoint-paper-bibfile}

\begin{thebibliography}{}

\bibitem[Aloui et~al., 2011]{aloui2011global}
Aloui, R., A{\"\i}ssa, M. S.~B., and Nguyen, D.~K. (2011).
\newblock Global financial crisis, extreme interdependences, and contagion
  effects: The role of economic structure?
\newblock {\em Journal of Banking \& Finance}, 35(1):130--141.

\bibitem[Aue et~al., 2009]{aue2009break}
Aue, A., H{\"o}rmann, S., Horv{\'a}th, L., Reimherr, M., et~al. (2009).
\newblock Break detection in the covariance structure of multivariate time
  series models.
\newblock {\em The Annals of Statistics}, 37(6B):4046--4087.

\bibitem[Avanesov et~al., 2018]{avanesov2018change}
Avanesov, V., Buzun, N., et~al. (2018).
\newblock Change-point detection in high-dimensional covariance structure.
\newblock {\em Electronic Journal of Statistics}, 12(2):3254--3294.

\bibitem[Banerjee et~al., 2019]{banerjee2019spectral}
Banerjee, S., Akbani, R., and Baladandayuthapani, V. (2019).
\newblock Spectral clustering via sparse graph structure learning with
  application to proteomic signaling networks in cancer.
\newblock {\em Computational Statistics \& Data Analysis}, 132:46--69.

\bibitem[Barnett and Onnela, 2016]{barnett2016change}
Barnett, I. and Onnela, J.-P. (2016).
\newblock Change point detection in correlation networks.
\newblock {\em Scientific reports}, 6:18893.

\bibitem[Basseville et~al., 1993]{basseville1993detection}
Basseville, M., Nikiforov, I.~V., et~al. (1993).
\newblock {\em Detection of abrupt changes: theory and application}, volume
  104.
\newblock Prentice Hall Englewood Cliffs.

\bibitem[Bekaert et~al., 2014]{bekaert2014global}
Bekaert, G., Ehrmann, M., Fratzscher, M., and Mehl, A. (2014).
\newblock The global crisis and equity market contagion.
\newblock {\em The Journal of Finance}, 69(6):2597--2649.

\bibitem[Bekaert et~al., 2011]{bekaert2011global}
Bekaert, G., Ehrmann, M., Fratzscher, M., and Mehl, A.~J. (2011).
\newblock Global crises and equity market contagion.
\newblock Technical report, National Bureau of Economic Research.

\bibitem[Bertero and Mayer, 1990]{bertero1990structure}
Bertero, E. and Mayer, C. (1990).
\newblock Structure and performance: Global interdependence of stock markets
  around the crash of october 1987∗.
\newblock {\em European Economic Review}, 34(6):1155--1180.

\bibitem[Bhattacharjee et~al., 2017]{bhattacharjee2017network}
Bhattacharjee, B., Shafi, M., and Acharjee, A. (2017).
\newblock Network mining based elucidation of the dynamics of cross-market
  clustering and connectedness in asian region: An mst and hierarchical
  clustering approach.
\newblock {\em Journal of King Saud University-Computer and Information
  Sciences}.

\bibitem[Boubaker et~al., 2016]{boubaker2016financial}
Boubaker, S., Jouini, J., and Lahiani, A. (2016).
\newblock Financial contagion between the us and selected developed and
  emerging countries: The case of the subprime crisis.
\newblock {\em The Quarterly Review of Economics and Finance}, 61:14--28.

\bibitem[Gallegati, 2012]{gallegati2012wavelet}
Gallegati, M. (2012).
\newblock A wavelet-based approach to test for financial market contagion.
\newblock {\em Computational Statistics \& Data Analysis}, 56(11):3491--3497.

\bibitem[Jin and An, 2016]{jin2016global}
Jin, X. and An, X. (2016).
\newblock Global financial crisis and emerging stock market contagion: A
  volatility impulse response function approach.
\newblock {\em Research in International Business and Finance}, 36:179--195.

\bibitem[Kau{\^e} Dal’Maso~Peron et~al., 2012]{kaue2012structure}
Kau{\^e} Dal’Maso~Peron, T., da~Fontoura~Costa, L., and Rodrigues, F.~A.
  (2012).
\newblock The structure and resilience of financial market networks.
\newblock {\em Chaos: An Interdisciplinary Journal of Nonlinear Science},
  22(1):013117.

\bibitem[Kazemilari and Djauhari, 2015]{kazemilari2015correlation}
Kazemilari, M. and Djauhari, M.~A. (2015).
\newblock Correlation network analysis for multi-dimensional data in stocks
  market.
\newblock {\em Physica A: Statistical Mechanics and its Applications},
  429:62--75.

\bibitem[Kenourgios et~al., 2007]{kenourgios2007financial}
Kenourgios, D., Samitas, A., and Paltalidis, N. (2007).
\newblock Financial crises and contagion: evidence for bric stock markets.
\newblock In {\em EFMA Vienna Meetings}.

\bibitem[Keshavarz et~al., 2018]{keshavarz2018sequential}
Keshavarz, H., Michailidis, G., and Atchade, Y. (2018).
\newblock Sequential change-point detection in high-dimensional gaussian
  graphical models.
\newblock {\em arXiv preprint arXiv:1806.07870}.

\bibitem[Kolar and Xing, 2011]{kolar2011time}
Kolar, M. and Xing, E. (2011).
\newblock On time varying undirected graphs.
\newblock In {\em Proceedings of the Fourteenth International Conference on
  Artificial Intelligence and Statistics}, pages 407--415.

\bibitem[Kolar and Xing, 2012]{kolar2012estimating}
Kolar, M. and Xing, E.~P. (2012).
\newblock Estimating networks with jumps.
\newblock {\em Electronic journal of statistics}, 6:2069 -- 2106.

\bibitem[Longstaff, 2010]{longstaff2010subprime}
Longstaff, F.~A. (2010).
\newblock The subprime credit crisis and contagion in financial markets.
\newblock {\em Journal of financial economics}, 97(3):436--450.

\bibitem[Lowry et~al., 1992]{lowry1992multivariate}
Lowry, C.~A., Woodall, W.~H., Champ, C.~W., and Rigdon, S.~E. (1992).
\newblock A multivariate exponentially weighted moving average control chart.
\newblock {\em Technometrics}, 34(1):46--53.

\bibitem[Luchtenberg and Vu, 2015]{luchtenberg20152008}
Luchtenberg, K.~F. and Vu, Q.~V. (2015).
\newblock The 2008 financial crisis: Stock market contagion and its
  determinants.
\newblock {\em Research in International Business and Finance}, 33:178--203.

\bibitem[Mantegna, 1999]{mantegna1999hierarchical}
Mantegna, R.~N. (1999).
\newblock Hierarchical structure in financial markets.
\newblock {\em The European Physical Journal B-Condensed Matter and Complex
  Systems}, 11(1):193--197.

\bibitem[Markwat et~al., 2009]{markwat2009contagion}
Markwat, T., Kole, E., and Van~Dijk, D. (2009).
\newblock Contagion as a domino effect in global stock markets.
\newblock {\em Journal of Banking \& Finance}, 33(11):1996--2012.

\bibitem[McCulloh and Carley, 2011]{mcculloh2011detecting}
McCulloh, I. and Carley, K.~M. (2011).
\newblock Detecting change in longitudinal social networks.
\newblock Technical report, Military Academy West Point NY Network Science
  Center (NSC).

\bibitem[Mizuno et~al., 2006]{mizuno2006correlation}
Mizuno, T., Takayasu, H., and Takayasu, M. (2006).
\newblock Correlation networks among currencies.
\newblock {\em Physica A: Statistical Mechanics and its Applications},
  364:336--342.

\bibitem[Namaki et~al., 2011]{namaki2011network}
Namaki, A., Shirazi, A., Raei, R., and Jafari, G. (2011).
\newblock Network analysis of a financial market based on genuine correlation
  and threshold method.
\newblock {\em Physica A: Statistical Mechanics and its Applications},
  390(21-22):3835--3841.

\bibitem[Nobi et~al., 2014]{nobi2014correlation}
Nobi, A., Lee, S., Kim, D.~H., and Lee, J.~W. (2014).
\newblock Correlation and network topologies in global and local stock indices.
\newblock {\em Physics Letters A}, 378(34):2482--2489.

\bibitem[Onnela et~al., 2003]{onnela2003dynamic}
Onnela, J.-P., Chakraborti, A., Kaski, K., and Kertesz, J. (2003).
\newblock Dynamic asset trees and black monday.
\newblock {\em Physica A: Statistical Mechanics and its Applications},
  324(1-2):247--252.

\bibitem[Park and Sohn, 2019]{park2018detecting}
Park, J.~H. and Sohn, Y. (2019).
\newblock Detecting structural changes in longitudinal network data.
\newblock {\em Bayesian Analysis}.

\bibitem[Peel and Clauset, 2015]{peel2015detecting}
Peel, L. and Clauset, A. (2015).
\newblock Detecting change points in the large-scale structure of evolving
  networks.
\newblock In {\em Twenty-Ninth AAAI Conference on Artificial Intelligence}.

\bibitem[Roy et~al., 2017]{roy2017change}
Roy, S., Atchad{\'e}, Y., and Michailidis, G. (2017).
\newblock Change point estimation in high dimensional markov random-field
  models.
\newblock {\em Journal of the Royal Statistical Society: Series B (Statistical
  Methodology)}, 79(4):1187--1206.

\bibitem[Wang et~al., 2018]{wang2018correlation}
Wang, G.-J., Xie, C., and Stanley, H.~E. (2018).
\newblock Correlation structure and evolution of world stock markets: Evidence
  from pearson and partial correlation-based networks.
\newblock {\em Computational Economics}, 51(3):607--635.

\bibitem[Zalesky et~al., 2012]{zalesky2012use}
Zalesky, A., Fornito, A., and Bullmore, E. (2012).
\newblock On the use of correlation as a measure of network connectivity.
\newblock {\em Neuroimage}, 60(4):2096--2106.

\bibitem[Zamba and Hawkins, 2006]{zamba2006multivariate}
Zamba, K. and Hawkins, D.~M. (2006).
\newblock A multivariate change-point model for statistical process control.
\newblock {\em Technometrics}, 48(4):539--549.

\bibitem[Zhao et~al., 2016]{zhao2016structure}
Zhao, L., Li, W., and Cai, X. (2016).
\newblock Structure and dynamics of stock market in times of crisis.
\newblock {\em Physics Letters A}, 380(5-6):654--666.

\bibitem[Zhou et~al., 2010]{zhou2010time}
Zhou, S., Lafferty, J., and Wasserman, L. (2010).
\newblock Time varying undirected graphs.
\newblock {\em Machine Learning}, 80(2-3):295--319.

\end{thebibliography}

\end{document}